\documentclass[10pt,twocolumn]{article}

\usepackage{arxiv}
\usepackage{times}
\usepackage{epsfig}
\usepackage{graphicx}
\usepackage{amsmath}
\usepackage{amssymb}
\usepackage[pagebackref=true,breaklinks=true,letterpaper=true,colorlinks,bookmarks=false]{hyperref}
\usepackage{booktabs}
\usepackage{multirow, makecell}

\usepackage{flushend}
\usepackage{ctable}

\usepackage{soul}
\usepackage{amsmath}
\usepackage{amssymb}
\usepackage{microtype}
\usepackage{wrapfig}

\def\figurePath{arxivimages/}
\def\myfigure#1#2{\begin{figure}[htb]\centering\includegraphics*[width = \linewidth]{\figurePath#1}\vspace{0cm}\caption{#2}\vspace{-0.25cm}\label{fig:#1}\end{figure}}
\def\mycfigure#1#2{\begin{figure*}[t]\centering\includegraphics*[clip, width = \linewidth]{\figurePath#1}\vspace{0cm}\caption{#2}\vspace{-0.25cm}\label{fig:#1}\end{figure*}}
\newcommand{\mywrapfigure}[3]{%
\begin{wrapfigure}{r}{#2\columnwidth}%
  \begin{center}%
    \includegraphics[width=#2\columnwidth]{\figurePath#1}%
    \vspace{0cm}%
    \caption{#3}%
    \label{fig:#1}%
  \end{center}%
\end{wrapfigure}%
\leavevmode%
}

\newcommand{\argmin}[1]{\underset{#1}{\operatorname{arg\,min}\ }}

\newcommand{\refSec}[1]{Sec.~\ref{sec:#1}}
\newcommand{\refFig}[1]{Fig.~\ref{fig:#1}}
\newcommand{\refEq}[1]{Eq.~\ref{eq:#1}}
\newcommand{\refTbl}[1]{Tbl.~\ref{tbl:#1}}

\soulregister\ref7
\soulregister\cite7
\soulregister\refFig7
\soulregister\cite7
\soulregister\ref7
\soulregister\pageref7
\soulregister\shortcite7
\soulregister\eg7
\soulregister\ie7
\soulregister\etal7
\soulregister\reftbl7

\DeclareGraphicsExtensions{.png,.jpg,.pdf,.ai,.psd}
\newcommand{\mysection}[2]{\section{#1}\label{sec:#2}}
\newcommand{\mysubsection}[2]{\subsection{#1}\label{sec:#2}}
\newcommand{\mysubsubsection}[2]{\subsubsection{#1}\label{sec:#2}}

\usepackage{pifont}
\newcommand{\cmark}{\checkmark}%
\newcommand{\xmark}{\scalebox{0.85}{\ding{53}}}%
\usepackage{adjustbox}
\newcolumntype{R}[2]{%
    >{\adjustbox{angle=#1,lap=\width-(#2)}\bgroup}%
    l%
    <{\egroup}%
}\newcommand*\rot[2]{\multicolumn{1}{R{#1}{#2}}}%

\begin{document}

\title{Neural View-Interpolation for Sparse Light Field Video}

\author{
    \parbox{\textwidth}{
        \centering
        Mojtaba Bemana$^1$\ \ \ 		 
        Karol Myszkowski$^1$\ \ \ 
        Hans-Peter Seidel$^1$\ \ \ 
        Tobias Ritschel$^2$
    }
    \\
    \\
    \parbox{\textwidth}{
        \centering
        $^1$MPI Informatik\qquad
        $^2$University College London
    }
}

\pagestyle{plain}

\renewcommand{\paragraph}[1]{{\textbf{#1}}}
\newcommand{\pipeline}{f}
\newcommand{\lightfield}{L}
\newcommand{\lightfields}{\mathcal L}
\newcommand{\lfSlice}{I}
\newcommand{\lfSlices}{\mathcal I}
\newcommand{\lfCoord}{\mathbf x}
\newcommand{\lfCoords}{\mathcal X}
\newcommand{\numberOfPixels}{{n_\mathrm p}}
\newcommand{\parameters}{\theta}
\newcommand{\lightField}{L}
\newcommand{\expected}{\mathbb E}
\newcommand{\sparseLFCoords}{\mathcal Y}
\newcommand{\sparseLFCoord}{\mathbf y}
\newcommand{\heldOutLFCoords}{\mathcal H}
\newcommand{\heldOutLFCoord}{\mathbf h}
\newcommand{\warpCoord}{\mathbf w}
\newcommand{\decode}{\mathtt{geom}}
\newcommand{\warp}{\mathtt{warp}}
\newcommand{\softOcc}{\mathtt{occ}}
\newcommand{\spatialNeighbor}{{\mathcal N_\mathrm s}}
\newcommand{\temporalNeighbor}{{\mathcal N_\mathrm t}}
\newcommand{\spaceTime}{\Omega}
\newcommand{\spaceTimeMap}{\omega}
\newcommand{\interpolatedLFCoord}{{\lfCoord'}}

\maketitle

\begin{abstract}
We suggest representing light field (LF) videos as ``one-off'' neural networks (NN), \ie a learned mapping from view-plus-time coordinates to high-resolution color values, trained on sparse views.
Initially, this sounds like a bad idea for three main reasons:
First, a NN LF will likely have less quality than a same-sized pixel basis representation.
Second, only few training data, \eg 9 exemplars per frame are available for sparse LF videos.
Third, there is no generalization across LFs, but across view and time instead.
Consequently, a network needs to be trained for each LF video. 

Surprisingly, these problems can turn into substantial advantages: 
Other than the linear pixel basis, a NN has to come up with a compact, non-linear \ie more intelligent, explanation of color, conditioned on the sparse view and time coordinates.
As observed for many NN however, this representation now is \emph{interpolatable}: if the image output for sparse view coordinates is plausible, it is for all intermediate, continuous coordinates as well.
Our specific network architecture involves a differentiable occlusion-aware warping step, which leads to a compact set of trainable parameters and consequently fast learning and fast execution.
\end{abstract}

\mysection{Introduction}{Introduction}
Light field (LF) video 
provides a complete visual representation of a dynamic scene.
Regrettably, this capability results in excessive storage, capture and processing requirements.
The redundancy in such data appears to be high -- but what is the right way of exploiting it?
We will here demonstrate, how a neural network (NN), involving the right differentiable rendering steps, becomes an compact and interpolatable representation of a LF video.

In particular, we investigate LF video that is sparse. 
Sparsity in the angular domain means capture from a practical rig of 3$\times$3 cameras instead of hundreds of observations in  dense LF.
This reduces the amount of data, but introduces the new challenge of interpolation.
The same holds in the temporal domain: frame rate can be reduced, but only if additional temporal interpolation is applied.
Ultimately, spatial and temporal sparsity can be combined, requiring even more advanced interpolation.
Such high-quality, high-speed interpolation is the challenge addressed in this article.

The industry solution to interpolation is streaming sparse images and estimating depth and using warping \cite{mark1997post}.
While NNs have been suggested to estimate depth or interpolate, we here, for the first time, suggest representing the entire LF as a NN.
This representation is a mapping from view angle and time (three dimensions) to pixel appearance in a high-resolution image  (millions of dimensions).
We train our NN on very sparse data, \eg 3$\times$3 images.
We find a NN that has come up with a compact, geometrically meaningful non-linear explanation of all observations will also produce suitable non-observed, \ie interpolated, views.
This ``interpolating effect'' has frequently been observed for NNs optimized for latent encoding-decoding \eg for faces.
For our task, a well-defined space (view and time) is readily available and the only requirement is to find the right non-linear decoding to benefit from interpolation.
Key to making this work is the right network structure, involving differentiable warping. 

The resulting method can learn to represent a LF in a NN in little time and decode it at high frame rates (ca. 20~Hz) and high resolution (1024$\times$1024) for arbitrary continuous view and time coordinates.
We compare the resulting quality to several other baselines (NN and classic) as well as to ablations of our approach.

\mysection{Previous Work}{PreviousWork}
Our work is rooted in LF and image-based rendering (IBR), it is inspired by general interpolation of 3D information as well as making use of differentiable rendering, which we all review now:

\paragraph{Lightfield view interpolation}
Levoy and Hanrahan \cite{levoy1996light} and Gortler \etal \cite{gortler1996lumigraph} were first to formalize the concept of a light field and to devise hardware to capture it.
An important distinction is that LFs can either be dense or sparse.
This is less defined by the number of images, but more by the distance between the views.
In this work we focus on wider baselines, with typically $M \times N$ cameras spaced by 5--10\,cm \cite{flynn2019deepview}, respectively a large disparity ranging up to 250 pixels \cite{dabala2016lightfield,Mildenhall2019}, where $M$ and $N$ are single-digit numbers, \eg 3$\times$3 or 5$\times$5.

\begin{table}[h]
\setlength{\tabcolsep}{0.4pt}
    \centering
    \begin{tabular}{lllllllllll}
         Method&
         &
         \rot{90}{1em}{View}&
         \rot{90}{1em}{Time}&
         \rot{90}{1em}{Sparse}&
         \rot{90}{1em}{Warp}&
         \rot{90}{1em}{Neural}&
         \rot{90}{1em}{One-off}&
         \rot{90}{1em}{Real-time}&
         \rot{90}{1em}{High-res}
         \\
         \toprule
         ULR&\cite{Buehler2001}&
         \cmark\hphantom{$^11$}& \cmark\hphantom{$^11$}& \cmark\hphantom{$^11$}& \cmark\hphantom{$^11$}& \xmark\hphantom{$^11$}& \xmark\hphantom{$^11$}& \cmark\hphantom{$^11$}& \cmark\hphantom{$^11$}\\
         Soft 3D&\cite{Soft3DReconstruction}&
         \cmark\hphantom{$^1$}& \cmark\hphantom{$^1$}& \cmark\hphantom{$^1$}& \cmark\hphantom{$^1$}& \xmark\hphantom{$^1$}& \xmark\hphantom{$^1$}& \cmark\hphantom{$^1$}& \cmark\hphantom{$^1$}\\
         Deep Blending& \cite{hedman2018blending}&
         \cmark& \cmark& \cmark& \cmark& \cmark\tmark[1]& \xmark& \cmark& \cmark\\
         Puppet Dubbing&\cite{ohad2019dubbing}& 
         \xmark& \cmark& \xmark& \xmark& \cmark\tmark[2]& \cmark& \xmark& \cmark\\
         Video-to-video & \cite{wang2018vid2vid}&
         \xmark& \cmark& \xmark& \xmark& \cmark\tmark[3]& \xmark& \xmark& \cmark\\
         Kalantari et al.&\cite{Kalantari2016}&
         \cmark& \xmark& \xmark& \cmark& \cmark\tmark[4]& \xmark& \xmark& \xmark\\
         Local LF Fusion &\cite{Mildenhall2019}&
         \cmark& \xmark& \cmark& \cmark& \cmark\tmark[5]& \xmark& \cmark& \xmark\\
         DeepView& \cite{flynn2019deepview}&
         \cmark& \xmark& \cmark& \xmark& \cmark\tmark[6]& \xmark & \xmark& \xmark\\
         Appearance Flow&\cite{zhou2016appearance}&
         \cmark& \xmark& \xmark& \cmark& \cmark\tmark[7]& \xmark& \cmark& \xmark\\
         DeepVoxels& \cite{sitzmann2019deepvoxels}&
         \cmark& \xmark& \xmark& \xmark& \cmark\tmark[8]& \cmark& \xmark& \xmark\tmark[10]\\
         Neural Volumes&\cite{lombardi2019volumes}&
         \cmark& \cmark& \cmark& \cmark\tmark[11]& \cmark\tmark[8]& \cmark&  \cmark& \xmark\tmark[10]\\
         Ours&&
         \cmark& \cmark& \cmark& \cmark& \cmark\tmark[9]& \cmark& \cmark& \cmark\\
         \bottomrule
    \end{tabular}
    \caption{Comparison of LF view synthesis-related methods.
    Learns:
    $^1$blend weights,
    $^2$appearance/audio labels,
    $^3$frame generation (conditional GAN),
    $^4$appearance,
    $^5$blend and opacity weights in output MPI,
    $^6$MPI gradient descent rules,
    $^7$warping,
    $^8$differential (tomographic) rendering,
    $^9$mapping camera coordinates to per-pixel scene geometry.
    Restricts: $^{10}$to motion compensation, not for rendering,
    $^{11}$128$\times$128$\times$128 voxels at most.
    }
    \label{tbl:PreviousWork}
\end{table}

For such sparse data, synthesizing intermediate views (interpolation or super-resolution) is an important problem that has received much attention as summarized in \refTbl{PreviousWork}.
The columns ``View'' and ``Time'' specify whether novel views can be derived in spatial and temporal domains, respectively.
``Sparse'' refers to the ability to handle LFs with wide baselines.
``Warp'' determines whether some form of explicit warping between neighboring views is performed.
``NN'' and ``One-off'' specify which methods are based on neural networks and if they need to be trained per scene, \ie they cannot generalize for other scenes but are specific to a particular scene.
The columns ``Real-time'' and ``High-res'' indicate that novel view rendering can be performed in real time (at least at 20~Hz) and at high-resolution (we aim for HD, \ie 1920$\times$1080).

A simple solution for interpolation is linear blending, but this leads to ghosting.
Unstructured lumigraph rendering (ULR) \cite{Buehler2001,Chaurasia2013} creates proxy geometry to warp \cite{mark1997post} multiple observations into a novel view and blend them with specific weights.
Recent work has used per-view geometry \cite{hedman2016scalablerendering} or NNs to compute the weights \cite{hedman2018blending}.
Our approach does not learn blending, but rather a deep representation of geometry itself that enables precise interpolation with occlusion handling.
Originally developed for unstructured sets of images, ULR-style IBR is a workable choice for LF video as well, in particular if analysis and novel-view-synthesis have to occur at real-time rates \cite{dabala2016lightfield}.
Avoiding the difficulty to reconstruct geometry has been addressed for LFs, without \cite{DuDDHM2013,Kellnhofer2017} or with \cite{zhou2016appearance} NNs.

An attractive recent idea is to learn synthesizing novel LF views.
One option is to explicitly computing a depth map \cite{Kalantari2016,Srinivasan2017} that explains the light field.
Our approach follows a similar route, but besides extension to time, we represent geometry as a NN, such that it becomes interpolatable.

Another option is to decompose the input LF into multiple depth planes of the output view and construct a view-dependent plane sweep volume (PSV) \cite{flynn2016deepstereo,Soft3DReconstruction,Xu2019}.
By learning how neighboring input views contribute to the output view, the multi-plane image (MPI) representation \cite{Zhou2018StereoMagnification} can be built that enables high-quality local LF fusion \cite{Mildenhall2019}.
Inferring a good MPI representation can be facilitated with learned gradient descent \cite{flynn2019deepview}, where the gradient components directly encode visibility and effectively inform the NN on the occlusion relations in the scene.
All these techniques avoid the problem of explicit depth reconstruction and allow for softer, and more pleasant results.
A drawback is the massive volumetric data, the difficulty to distribute occupancy in it, and finally volume rendering itself.

Other work has gone fully volumetric for arbitrary views. 
Deep Voxels \cite{sitzmann2019deepvoxels} in particular takes a high number of images and learns how to find a deep 3D representation that can be reprojected into many views.
Notably, this is a NN trained per scene (column ``One-off'' in \refTbl{PreviousWork}), but without exploiting the interpolation property.
Also, frequently \cite{nguyen2018rendernet}, the differentiable tomographic rendering step is learned, while in our approach, a differentiable warping with occlusion handling is used that does not require learning any parameters and can work with off-the-shelf warping.
Recent work has extended this into the time domain \cite{lombardi2019volumes}, and is closest to our approach.
They also use warping, but for a very different purpose: deforming a pixel-basis 3D representation over time in order to save storing individual frames (motion compensation).
Both methods \cite{sitzmann2019deepvoxels,lombardi2019volumes} are limited by the spatial 3D resolution of volume texture and the need to process it, while we work in 2D depth and color maps only.
Ultimately, we do not claim depth maps or volumes to be better or worse per-se, but would suggest that 3D volumes have their strength for seeing objects from all views (at the expense of resolution), whereas our work, using images, is more for observing scenes from a ``funnel'' of views, but at high 2D resolution.
No work yet is able to combine high resolution and arbitrary views, not to speak of time.

\paragraph{Interpolation} of sparse observations is an important computer graphics problem, ranging from a single pixel to a full LF and extends to many domains.
We have discussed LF interpolation above, but our work also is inspired by work in other domains.

Interpolating reflectance fields \cite{Fuchs2007} is a related problem, where related solutions have been suggested: Ren and colleagues use a simple one-off neural network for representation \cite{ren2015image}.
Rainer~et~al.~\cite{rainer2019btf} have used more modern encoding-decoding to compress spatially-varying reflectance.
Maximov~et~al.~\cite{maximov2019dam} encode appearance (the combination of illumination and reflectance) as a NN.
In all these works, observations are spatially registered and generalization is across view or light with no challenges of space-time geometry.
In this work we deal with appearance that changes across space and time.
Videos, as LF videos, comprise of discrete frames.
To get smooth interpolation, \eg for slow-motion (individual frames), motion blur (averaging multiple frames) images need to be interpolated, potentially using NNs \cite{vondrick2016generating,liu2017video,Sun2018PWC-Net}. 
More exotic domains of video re-timing, which involve annotation of a fraction of frames and one-off NN training, include the space of visual in sync to spoken language \cite{ohad2019dubbing}.
Even more extreme is temporally-consistent video content generation using conditional GANs \cite{wang2018vid2vid}.

A key inspiration for this work is the coord-conv trick \cite{liu2018coordconv}.
Their didactic examples show, how a NN in conjunction with their trick, has the ability to make sense of a very limited set of images to a level that it can fill the gaps faithfully, \ie interpolate with high plausibility.
While their paper shows  single moving white square on a black background depending on a pair of coordinates to control it, we apply it to real visual data as complex as LF videos, depending on angle and time.

\paragraph{Differentiable rendering}
To learn geometric structure from observations with no direct supervision, differentiable rendering has become popular.
The MonoDepth \cite{monodepth17}, system is an excellent example: Here a network learns to regress disparity for pairs of images such that each image in the pair can explain the other.
This does not require supervision by depth.
We follow the same idea, but extend this to occlusion handling, learning the combination and representing depth itself as a network for interpolation.
MonoDepth among others uses the Spatial Transform Layer \cite{jaderberg2015spatial} to warp one stereo pair image to the other view.

Handling of occlusion is an important computer graphics problem and recently several methods were proposed to include it in a differentiable pipeline.
The typical solution is to smooth binary occlusion \cite{liu2019softras,petersen2019pix2vec}, which is also what we do. 
In the same vein, for synthesizing appearance from other views, it has been shown \cite{zhou2016appearance} that regressing the transformation is superior to regressing appearance itself.
Our work extends this to space time and combines it with handling of occlusion.

\mysection{Background}{Background}

Two main observations motivate our approach: First, representing information using NNs leads to interpolation.
Second, this property is retained, if the network contains more useful layers, such as a differentiable rendering step.
Both will be discussed next:

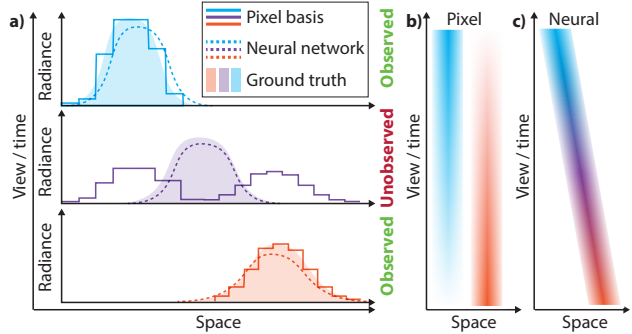
\myfigure{NNInterpolation}{NN and pixel interpolation:
\textbf{a)} Flatland interpolation in the pixel \textbf{(lines)} and the NN representations \textbf{(dotted lines)} compared to a reference \textbf{(solid)} for a 1D light field (vertical axis angle; horizontal axis space).
The top and bottom are observed and the middle is unobserved \ie interpolated.
\textbf{b,c)} Comparing the continuous interpolation in the pixel and the NN representation visualized as a (generalized) epi-polar image.
Note that the NN leads to smooth interpolation, while the pixel representation causes undesired fade-in/fade-out transitions.
}
\paragraph{Deep representations help interpolation.}
It is well-known, that deep representations amend to interpolation of 2D images \cite{radford2015unsupervised,reed2015deep,white2016sampling}, audio \cite{engel2017neural} or 3D shape \cite{dosovitskiy2015learning} much better than the pixel basis.

Consider the blue and orange bumps in \refFig{NNInterpolation}, a; these are observed.
They represent flat-land functions of appearance (vertical axis), depending on some abstract domain (horizontal axis), that later will become space, or time or both in out LF video problem.
We wish to interpolate something close to the unobserved violet bump in the middle.
Linear interpolation in the pixel basis (solid lines), will fade both in, resulting in two flat copies.
Visually this would be unappealing and distracting ghosting.
This difference is also seen in the continuous setting of \refFig{NNInterpolation}, b that can be compared to the reference in \refFig{NNInterpolation}, c.
When representing the bumps as NNs to map coordinates to color (dotted lines), we note:
They are slightly worse than the pixel basis and might not match the NNs.
However, the interpolated, unobserved result is much closer to the reference, and this is what matters in LF video interpolation.

Typically, substantial effort is made to construct encoding into and decoding from these deep representations such as with auto-encoders \cite{hinton2006reducing}, variational auto-encoders  \cite{kingma2013auto} or adversarial networks \cite{goodfellow2014generative}.
In our problem we already have the latent space given as beautifully laid-out space-time LF coordinates and only need to learn to decode these into images.

\paragraph{(Differentiable) rendering is just another non-linearity.}

The second key insight is, that the above property is not affected by inserting more advanced layers such as differentiable rendering (warping) into a learnable pipeline.
\refFig{NNInterpolation} shows interpolation of colors over space.
We find the interpolation property to be retained, if the NNs is made more-fit-for purpose, than, say, vanilla regression of appearance using a multi-layer perceptron (MLP) or convolutional neural network (CNN).
\mywrapfigure{Zebra}{0.37}{Zebra space}
CNNs without the cord-conv \cite{liu2018coordconv} trick are particularly bad at such spatially-conditioned generation.
But even with coord-conv, this complex function is unnecessarily hard to find and slow to fit.
In particular, consider comparing  such a CNN/MLP to a design that is able to sample the observations such as proposed by spatial transformers \cite{jaderberg2015spatial,zhou2016appearance}, a very primitive form of differentiable rendering.
Learning all stripes in \refFig{Zebra} is harder than learning how they move coherently when changing view.

We will now detail our work, motivated by those observations.
\mycfigure{Overview}{
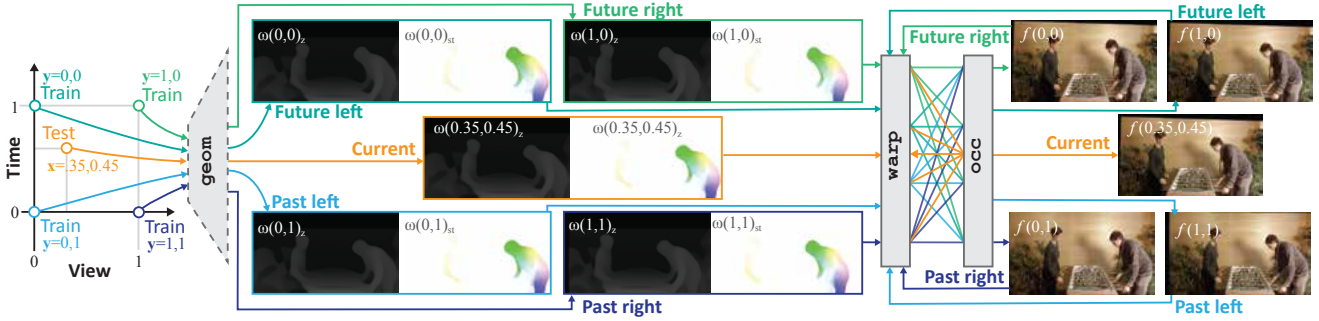
Overview of our approach, left to right:
For training, starting from space-time coordinates $x$ and $t$ (here 1D space), the learnable \texttt{geom} layer decodes the per-pixel space-time geometry map $\spaceTimeMap$.
This geometry map is input to a differentiable reprojection layer that \texttt{warp}s every observation into all other coordinates.
We here show the depth as grey and the temporal flow map as false-color visualizations.
An occlusion layer \texttt{occ} combines all observations with each other globally.
The loss minimizes the error between reprojection and observation.
During training, the observation in the corners of the space are fed to train the model (greenish colors).
We call this step a decoder, as it decodes LF coordinates into geometry maps. 
During test (orange colors), fractional coordinates are used, here the random point $(0.35, 0.45)$.
The only learnable mapping is \texttt{geom} denoted with a dotted line.
All other steps are differentiable but have no learnable parameters.}

\mysection{Our Approach}{OurApproach}

We will first give a formal definition of the function we learn, followed by the network architecture we choose for implementing it.

\mysubsection{Objective}{Objective}
We represent the light field video as a non-linear function $
\pipeline_\parameters(\lfCoord)
\in
\lfCoords
\rightarrow
\mathbb R^\numberOfPixels,
$
where $\lfCoord=(u,v,t)\in\lfCoords$ is the \emph{light field coordinate} (two spatial coordinates $u,v$ and a temporal dimension $t$) in the light field coordinate system $\lfCoords\subset\mathbb R^3$, 
and $\numberOfPixels$ is the number of pixels (millions).
Please note the different coordinate systems (two-plane parametrization of a LF): $\lfCoord$ allocates different images in space and time, not horizontal or vertical coordinates inside an image.

We denote as $\sparseLFCoords\subset\lfCoords$ the subset of \emph{observed} LF coordinates for which we know the \emph{light field image} denoted as  $\lightfield(\sparseLFCoord)$.
Typically $|\sparseLFCoords|$ is sparse, \ie small, like 3$\times$3 or 5$\times$5.
We find this mapping $\pipeline$ by optimizing for
\[
\parameters=
\argmin{\parameters'}
\expected_{\sparseLFCoord\sim\sparseLFCoords}
||
\pipeline_{\parameters'}(\sparseLFCoord)
\ominus
\lightfield(\sparseLFCoord)
||_1
\] where $\expected_{\sparseLFCoord\sim\sparseLFCoords}$ is the expected value across all the discrete and sparse LF coordinates $\sparseLFCoords$,
$\ominus$ is a perceptual image difference.
In prose, we train a NN $\pipeline$ to map 3D LF coordinates $\sparseLFCoord$ to images of observed samples $\lightfield(\sparseLFCoord)$ of the light field.

Note, that training never evaluates any LF coordinate $\lfCoord$ that is not in $\sparseLFCoords$, as we would not know what the image $\lightfield(\lfCoord)$ at that location is.
But as $\pipeline$ is an ``intelligent'' non-linear explanation for a few $\sparseLFCoord$ it generalizes from the discrete observed coordinates $\sparseLFCoords$  to the unobserved continuous $\lfCoords$.
Note, that as we aim for interpolation, $\lfCoords$ is a convex combination of $\sparseLFCoords$ and does not extend beyond.

\mysubsection{Architecture}{Architecture}
Our pipeline $\pipeline$, depicted in \refFig{Overview}, has three  main steps summarized in \refSec{Interpolation}:
representing space-time geometry using a NN denoted $\decode$ (\refSec{Decoder}),
warping $\warp$ according to that representation (\refSec{Warping}) and
resolving occlusion (\refSec{SoftOcclusion}) using a step $\softOcc$.

The system is implemented in TensorFlow and trained using ADAM optimizer with a learning rate of 0.0001.

\mysubsubsection{Interpolaion}{Interpolation}
We first resolve spatial interpolation (\refEq{Spatial}), followed by a temporal one (\refEq{Temporal}).
This choice is arbitrary but results in subtle differences.

Spatial interpolation creates an intermediate LF defined as
\begin{align}
\label{eq:Spatial}
\bar\lightfield
(\lfCoord)
=
\sum_{\sparseLFCoord\in
\spatialNeighbor(\lfCoord)}
\softOcc(\lfCoord, \sparseLFCoord)
\odot
\warp(\lightfield(\sparseLFCoord), \decode(\sparseLFCoord), \lfCoord)
,
\end{align}
where
$\spatialNeighbor(\lfCoord)$ is the set of all spatial neighbors of $\lfCoord$ and $\odot$ is per-pixel (Hadamard) multiplication. 
Interpolation sums all spatial neighbors $\spatialNeighbor(\lfCoord)$ but excludes the observation at $\lfCoord$ itself.
In \refFig{Overview}, different inputs to $\pipeline$ with different coordinates are encoded as colors.
Every observation (blue-green colors) has to explain itself using geometry from all others at training time.
In this example the 4 observations regularly cover the unit space-time  quad, with a single spatial dimension only.
At test time, the pipeline is ran with a continuous, in-between $\lfCoord$, denoted as an orange dot.

For each observation, three steps occur: $\decode$, $\warp$ and $\softOcc$.
Geometry at that LF coordinate $\sparseLFCoord$ is represented using a trainable unit $\decode$ (\refSec{Decoder}).
\refFig{Overview} shows the output of that unit in its center, both for training (blue-greenish) and for testing (orange) coordinates.
Using this geometry, the observation is warped to the desired unobserved light field coordinate $\lfCoord$ using $\warp$ (\refSec{Warping}).
\mywrapfigure{InterpolationScheme}{0.37}{Interpolation.}
Finally, the warped observation is weighted using a soft occlusion term $\softOcc$ that will give lower weights if the value required at $\lfCoord$ was occluded in $\sparseLFCoord$ (\refSec{SoftOcclusion}).
\refFig{Overview} shows dense links between warping and occlusion as all warped observations are resolved jointly.

Similar to the spatial one, temporal interpolation is 
\begin{align}
\label{eq:Temporal}
\pipeline(\lfCoord)
=
\sum_{\interpolatedLFCoord\in
\temporalNeighbor(\lfCoord)}
\softOcc(\lfCoord, \interpolatedLFCoord)
\odot
\warp(\bar\lightfield(\interpolatedLFCoord), \decode(\interpolatedLFCoord), \lfCoord)
,
\end{align}
where
$\temporalNeighbor(\lfCoord)$ are the temporal neighbors of $\lfCoord$ and
$\bar\lightfield(\interpolatedLFCoord)$ is the spatially interpolated lightfield resulting from \refEq{Spatial}.
At the temporal neighbors $\interpolatedLFCoord$ (\refFig{InterpolationScheme}), the light field $\lightfield(\interpolatedLFCoord)$ is not observed.
Hence, the spatial interpolation $\bar\lightfield$ is used as a proxy.
We denote coordinates into the already-interpolated LF as $\interpolatedLFCoord$.

We will detail all steps in the next sections \ref{sec:Decoder}--\ref{sec:SoftOcclusion}.

\mysubsubsection{Decoder}{Decoder}
Input to the decoder is the LF coordinate $\lfCoord$ and output is a per-pixel 
space-time geometry map \[
    \decode(\lfCoord)
    \in
    \lfCoords\rightarrow\spaceTime,
\]
where $\spaceTime=(0,1)^{\numberOfPixels\times 3}$.
This map has three channels for all $\numberOfPixels$ pixels $\spaceTimeMap\in\spaceTime$.
The first one $\spaceTimeMap_\mathrm z$ is related to space; a depth map.
The second and third component $\spaceTimeMap_\mathrm u$ and $\spaceTimeMap_\mathrm v$ are related to motion; a flow map.
As the camera transformations between views is known, this is sufficient to map from one view to another.
Temporal information is in units of 2D per-pixel motion and frames are regular in time.
The decoder could also be considered a conditional generator.
We detail use of this space-time geometry information further when explaining the details of warping (\refSec{Warping}).

Please also note, that no pixel-basis observation
$\lightfield(\sparseLFCoord)$ ever is input to the decoder, and hence, all geometric structure is encoded in the network.
Recalling \refSec{Background}, we see this is both a burden, but also required to achieve the desired  interpolation property: if the geometry NN can explain the observations at a few $\sparseLFCoord$, it can explain their interpolation at all $\lfCoord$.
This also justifies why $\decode$ is a NN and we do not directly learn a pixel-basis depth-motion map: it would not be interpolatable.

We found the particular details of $\decode$ to be less relevant.
Our implementation starts with a fully connected operation that transforms the coordinate into a 2$\times$2 image with 128 channels.
The coord-conv \cite{liu2018coordconv} information is added at that stage.
This is followed by as many steps as it takes to arrive at the output resolution, reducing the number of channels to arrive at 3 output channels in the end.
Note, that using skip connections is not applicable to our setting, as the decoder input is a mere three numbers without any spatial meaning.

\mysubsubsection{Warping}{Warping}
Warping is defined as the mapping
\[
    \warp(\lfSlice, \spaceTimeMap, \lfCoord)
    \in
    \lfSlices\times\spaceTime\times\lfCoords
    \rightarrow
    \lfSlices.
\]
from the LF slice $\lfSlice\in\lfSlices$ (an image),
the space-time geometry $\spaceTimeMap$ and
the unobserved LF coordinate $\lfCoord$ to an image.

Warping needs to interpret the geometry $\spaceTimeMap$ to get an idea where to sample \cite{jaderberg2015spatial} the observed image $I$.
This involves constructing an inverse warp, \ie finding which pixel in the observed $\lfSlice$ maps to each pixel in $\lightfield(\lfCoord)$.
Constructing such inverse warps is possible \cite{YangNehab2009,DidykERMS2010,bowles2012iterative,leimkuehler2017microwarping}, but requires operations that are difficult to back-propagate.
Instead, we make the simplifying assumption, that the inverse flow is the negation of the forward flow.
This assumptions holds for planar geometry \cite{bowles2012iterative}.
Different from warping where the depth is given (like a z-buffer from rasterization) our method optimizes over depth to please warping.
Now, learning will choose depth values, such that when inserted into the warping, will best explain the image.
This includes avoiding depth that causes difficulties to warping, \ie deviations from the model assumptions.
It could be said that here, data is fit to the model.

Constructing the forward flow is done differently for space and time.
In space, we use the known spatial arrangement (baseline, directions, etc.) to convert the first channel of the geometry into disparity.
In time, flow is assumed to be symmetric at the current frame, so finding backward motion from forward motion is simple negation.
These assumptions drastically reduce the degrees of freedom to one in space and two in time as well as they impose additional constraints that regularize the problem.

Note, that $\warp$, while differentiable, does not have any learnable parameters and is very effective deployment: a single bi-linear texture lookup.

\mysubsubsection{Soft occlusion implementation}{SoftOcclusion}
To combine an interpolation from an input LF coord $\lfCoord'$ with an output LF coord $\lfCoord$, we again make use of the geometry model learned: The model has to report depth of $\lfCoord$ to be smaller than $\lfCoord'$ for that point to be visible.
Such a hard decision however is not differentiable and introduces visually distracting discontinuities.
For those two reasons, we make use of a soft occlusion term, defined as
\[
\softOcc(\lfCoord, \interpolatedLFCoord)=
\frac{
\exp(
-
|
\decode(\lfCoord)_\mathrm z-
\decode(\interpolatedLFCoord)_\mathrm z
|
)
}{
\sum_i
\exp
(
-
|
\decode(\lfCoord)_\mathrm z-
\decode(\interpolatedLFCoord_i)_\mathrm z
|
)
}.
\]
In other words, depth values from observed images are down-weighted, when the position they are taken from indicates, they would not be similar to the pixel depth they will end up with.
The denominator makes sure the positive weights form a partition of unity by iterating all other $\interpolatedLFCoord_i$ contributing to occlusion handling.
Note, that this weighting is a global construction: the weight of one observation depends on all others as well as the output coordinate.

Simpler (hard) forms of occlusion consistency have been used in LF analysis \cite{WG13_tpami}, but we here, for the first time, inspired by differentiable soft-occlusion rendering \cite{liu2019softras,petersen2019pix2vec}, include time and use it in a trainable architecture.

\mysection{Results}{Results}
Here we will provide comparison to other work (\refSec{ComparisonResults}), evaluation of our scalability (\refSec{EvaluationResults}) as well as applications (\refSec{ApplicationResults}).

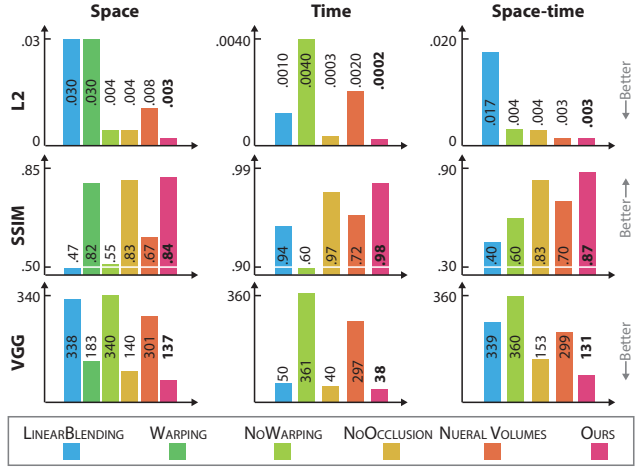
\myfigure{QuantitativeResults}{Results of different methods \textbf{(colors)}, on different domains \textbf{(columns)} according to three metrics \textbf{(rows)}.
Our method performs best in all tasks according to all metrics on all domains.}

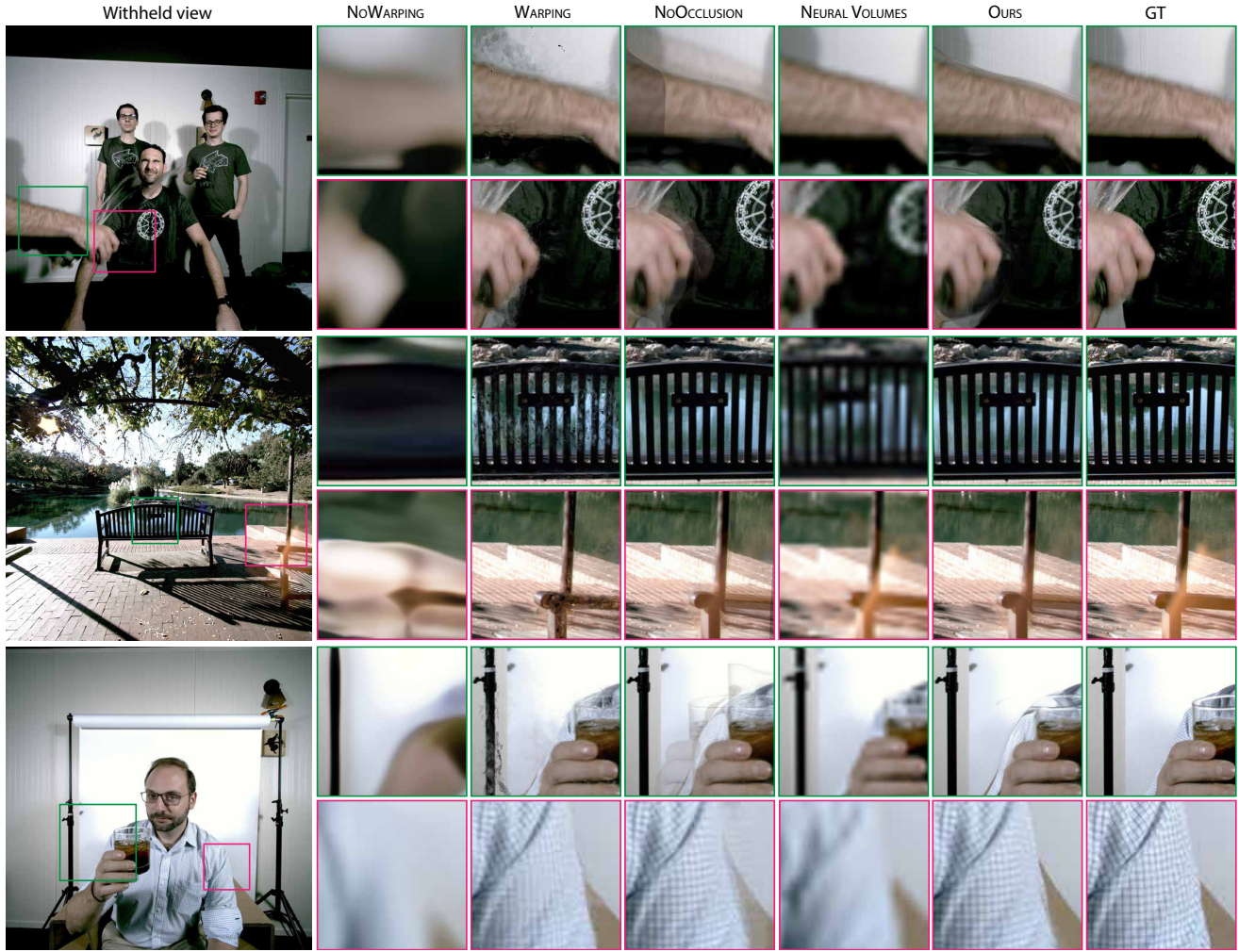
\mycfigure{SpatialComparison}{Comparison of our approach for view interpolation to other methods at three scenes \textbf{(rows)}.
Columns show, left to right, the withheld reference, the results by (\textsc{NoWarping}, traditional \textsc{Warping}, \textsc{NoOcclusion}, \textsc{NeuralVolumes}, and \textsc{Ours}), as well as the ground truth as insets.}

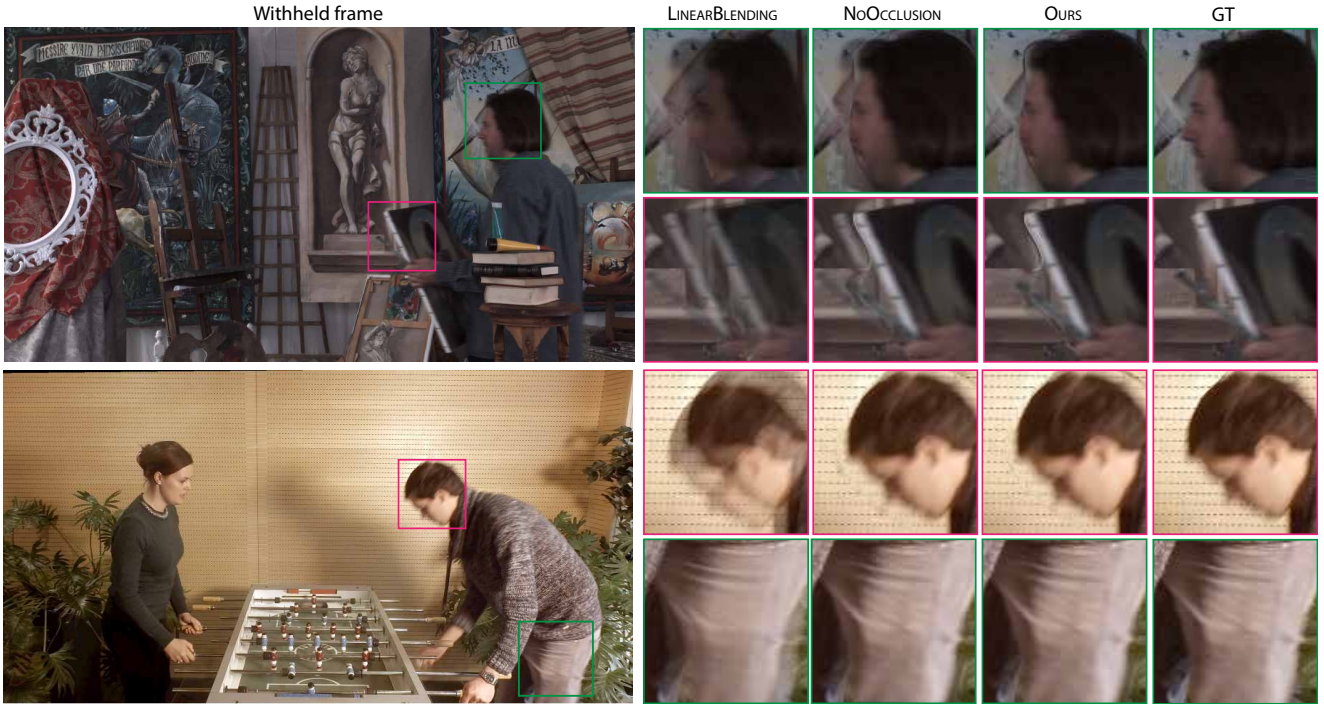
\mycfigure{TemporalComparison}{Temporal interpolation for two scenes \textbf{(rows)} using different methods \textbf{(columns)}.
Temporal resolution is $30$ frames.
Spatial resolution is 960$\times$540.
See \refSec{ComparisonResults} for a discussion.}

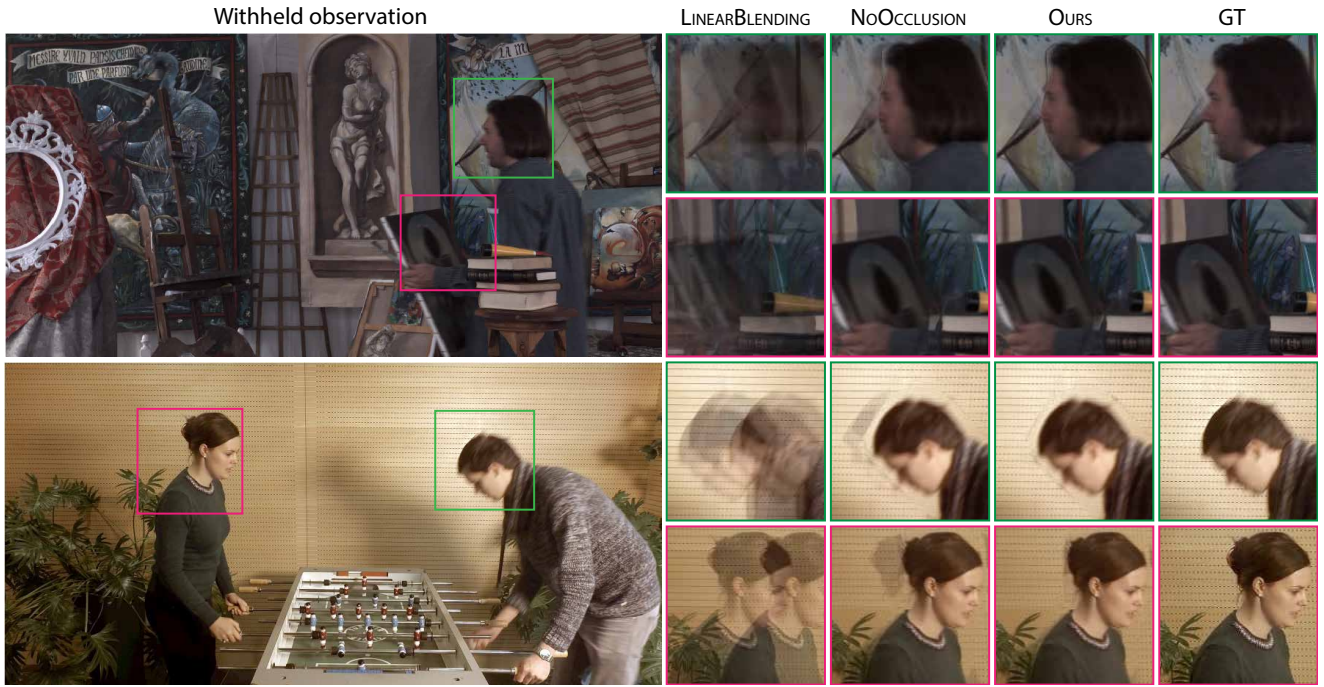
\mycfigure{SpaceTimeComparison}{Results for space-time interpolation.
View observations resolution is 3$\times$5.
Spatial resolution is 960$\times$540, and temporal resolution is $30$ frames.}

\mysubsection{Comparison}{ComparisonResults}
We compare our approach to other \emph{methods}, following a specific \emph{protocol} and by different \emph{metrics} to be explained now:

\paragraph{Methods}
Comparison is made against six methods: \textsc{Ours},
\textsc{Blending},
\textsc{Warping},
\textsc{NeuralVolume}, 
and two ablations
\textsc{NoWarping} and
\textsc{NoOcclusion}.

Linear \textsc{Blending} is not a serious method, but documents the sparsity: plagued by ghosting for small baselines, we see our baseline / sparsity poses a difficult interpolation task, far from linear.

\textsc{Warping} first estimates the depth using a light field video depth estimation method \cite{dabala2016lightfield} and later applies warping \cite{mark1997post} with ULR-style weights \cite{Buehler2001}.
These depth estimators include consistency voting, eliminate outliers, perform bilateral sampling etc. and can be considered an upper bound on what classic methodlogy is able to do today.
Note, how ULR weighting accounts for occlusion.

\textsc{NeuralVolumes} compares to recent NN-based view interpolation making use of volumes and ray-marching \cite{lombardi2019volumes} and \cite{sitzmann2019deepvoxels}.
We choose to compare to the more recent NeuralVolumes \cite{lombardi2019volumes}, as it also supports time.
It is assumed, that their method perfectly manages to create the ground-truth volume at the 3D resolution they used ($128^3$) and it was able to ray-march it without any error to produce a $128^2$ image.
So we simply down-sample the ground truth to 128 and upsample it again.

Finally, we compare two ablations of our method.
The first, \textsc{NoWarping} uses direct regression of color values without warping.
The second, \textsc{NoOcclusion} does not perform occlusion reasoning but averages directly.

\paragraph{Protocol}
Success is quantified as the expected ability of a method to predict a set of held-out LF observed coordinates $\heldOutLFCoords$ when trained on $\sparseLFCoords-\heldOutLFCoords$, \ie $\expected_{\heldOutLFCoord\sim\heldOutLFCoords}\pipeline(\heldOutLFCoord)\ominus_\mathrm m\lightfield(\heldOutLFCoord)$, where $\ominus_\mathrm m$ is one of the metrics to be defined below.
The held-out set can be a single or multiple observation and can be across space or time or both.

\paragraph{Metrics}
For comparing the predicted to the held out view we use $L_2$, DSSIM and VGG.
In all cases, less is better.
We also measure the joint time of pre-processing, if required.

\paragraph{Data}
We use the publicly available data from \cite{levoy1996light} and \cite{Soft3DReconstruction} for spatial light fields, and LF video data from \cite{dabala2016lightfield,Sabater2017}.

\paragraph{Results}
\refFig{QuantitativeResults} summarizes the outcome of the main comparison.
We see, that our method provides the best quality in all tasks according to all metrics on all domains.

For example images corresponding to the plots in \refFig{QuantitativeResults}, please see \refFig{SpatialComparison} for interpolation in space, \refFig{TemporalComparison} for time and \refFig{SpaceTimeComparison} for space-time results.
In each figure we document the input view and multiple insets that show the results by all competing methods.
Linear blending does not work and shows that views are substantially different and have complex disparity.
\textsc{NoWarping} can regress plausible colors, but without details.
\textsc{Warping} produces crisp images, but pixel-level outliers that are distracting in motion, \eg for the bench in \refFig{SpatialComparison}.
\textsc{NoOcclusion} results in crisper images but when multiple objects overlap it results in ghosting.
\textsc{NeuralVolumes} has cannot reproduce details, as seen in the shirt of the third line in \refFig{SpatialComparison}.
\textsc{Ours} has details, plausible motion and is generally most similar to the ground truth.
The temporal interpolation comparison in \refFig{TemporalComparison} indicate similar conclusions: \textsc{Blending} is no option, not handling occlusion, also in time, creates ghosting due to overlap.
We do not show \textsc{NeuralVolume} for the results to follow as it is a smooth version of our ground truth images with clear lack of details.
Ultimately, \textsc{Ours} is similar to the ground truth.
The motion smoothness is best seen in the slow-mo application of the supplemental video.
When interpolating across space and time as in \refFig{SpaceTimeComparison} ghosting effects get stronger for others, as images get increasingly different.
\textsc{Ours}, can have difficulties where deformations are not fully rigid as seen for faces, but overall compensates for this to produce plausible images.
Space-time results are also shown in the supplemental video.

We conclude, that both numerically and visually our approach can produce state-of the art interpolation in view and time in high spatial resolution and at high frame rates.
We will next look into evaluating the dependency of this success on different factors.

\mysubsection{Evaluation}{EvaluationResults}
Here we evaluate our approach in terms of scalability with training effort and observation sparsity, speed and detail reproduction.

\paragraph{Training effort}
Our approach needs to be trained again for every LF.
Typical training time is listed in \refTbl{TrainingTime}.
The results shown for time and space-time interpolation are one hour of training for a 3$\times$5 array of camera setup with resolution 960$\times$540 with 30 frames.

\begin{table}[h]
\setlength{\tabcolsep}{3pt}
    \centering
    \caption{Comparison of training time and network parameters for different resolutions for a 5$\times$5 LF array and spatial interpolation.}
    \begin{tabular}{llcccc}     
     &
     512$\times$512
     &
     1024$\times$1024
     &
     1764$\times$1764
     \\
     \toprule     
     Training time
     & 
     28 min. & 60  min.& 172  min.
     \\     
     Network parameters& 
     482,753& 492,001& 492,001\\    
     \bottomrule
\end{tabular}
\label{tbl:TrainingTime}
\end{table}

\refFig{Iteration} shows progression of interpolation quality over learning time.
We see, that even after little training, results can be acceptable, at least, better than all competitors after complete training.

Overall, we see that learning the NN requires  a workable amount of time, compared to the time other networks require that are in the order of many hours or days on many GPUs.

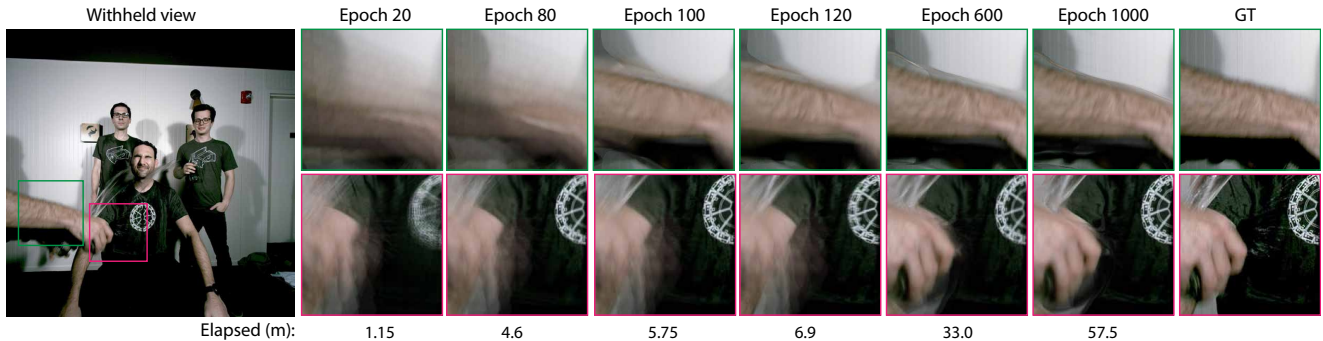
\mycfigure{Iteration}{Progression of visual fidelity for different training effort (horizontal) for two insets (vertical) in one scene.
After 500 epochs (ca.~30 minutes) the results is usable and converges after 1000 epochs (ca.~1h).
Note, that epochs are short as we only have 5$\times$5 training examples.}

\paragraph{Observation sparsity}
We interpolate form extremely sparse data.
In \refTbl{SparsityEvaluation} we have evaluate the quality of out interpolation depending on the number of training exemplars.
A visual representation of that progression is seen in \refFig{Sparsity}.

\begin{table}[h]
\setlength{\tabcolsep}{3pt}
    \centering
    \begin{tabular}{crrr}         
         \multicolumn{1}{c}{LF}&
         \multicolumn{1}{c}{VGG19}&
         \multicolumn{1}{c}{L2}&
         \multicolumn{1}{c}{SSIM}\\
         \toprule
          3$\times$3&  261& .010& .66\\\
          5$\times$5&  205& .005& .82\\
          9$\times$9&  148& .003& .89\\
          \bottomrule
    \end{tabular}
    \caption{Reconstruction error for the \emph{Crystal Ball} scene with resolution 512$\times$512 using different metrics \textbf{(columns)} for different angular density \textbf{(rows).}}
    \label{tbl:SparsityEvaluation}
\end{table}

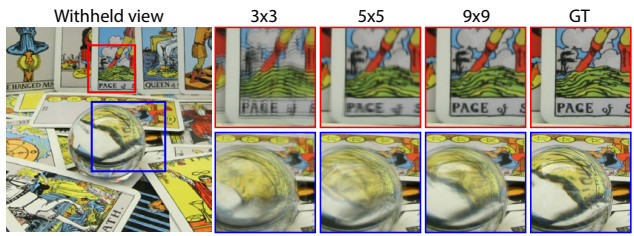
\myfigure{Sparsity}{Visual quality of our approach as a function of increasing \textbf{(left to right)} training set-size for angular interpolation.}

\paragraph{Speed}
At deployment, our method requires no more than taking three numbers and pass them through a decoder for each observation, followed by warping and a weighting.
The end-speed is around 20~Hz (on average 46\,ms per frame) at 1024$\times$1024 for a 5$\times$5 LF on a Nvidia 1080Ti  with 12 GB RAM.
Most of the time (31\,ms) is consumed in the non-optimized warping step in TensorFlow, that could be made much faster with an OpenGL implementation \cite{DidykERMS2010,bowles2012iterative}.

\paragraph{Smoothness}
The depth and flow map we produce are smooth in space and time and may lack detail.
\mywrapfigure{Smoothness}{0.5}{Smooth geometry.}
It would be easy to add skip connections to get the details form the appearance.
Regrettable, this would only work on the input image, and this needs to be withheld at training, and is unknown at test time.
An example of this smoothness if seen in \refFig{Smoothness}.
This smoothness is a main source of artifacts.
Overcoming this, \eg, using an adversarial design, is left to future work.

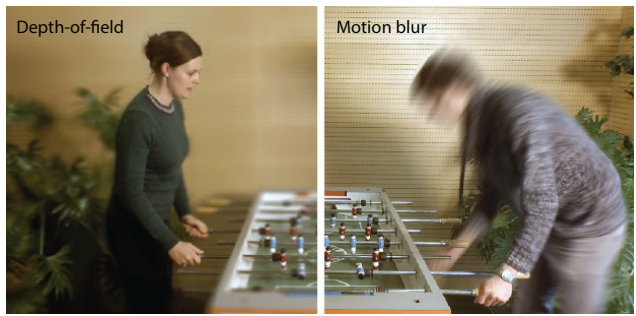
\myfigure{Application}{Two LF video-enabled effects, computed using view interpolation: Depth-of-field \textbf{(left)} and motion blur \textbf{(right)}.}

\mysubsection{Applications}{ApplicationResults}
\refFig{Application} demonstrates motion blur (interpolation across time) and depth-of-field (interpolation across angle), and both (interpolation across space and angle).


\mysection{Conclusion}{Conclusion}
We have demonstrated that representing a lightfield video as a NN that produces images conditions on view and time leads to high-quality, high-performance interpolation in space and time.
The particular structure of a network that combines a learnable space-time geometry model, combined with warping and reasoning on occlusion, has shown to perform better than direct regression of color or warping without handling occlusion.
In future work, we aim to further reduce training time (eventually using learned gradient descent \cite{flynn2019deepview}), explore interpolation along other domains such as illumination, wavelength or spatial audio \cite{engel2017neural}, and reconstruction from even sparser observations. 


\begin{thebibliography}{10}\itemsep=-1pt

\bibitem{bowles2012iterative}
H.~Bowles, K.~Mitchell, R.~W. Sumner, J.~Moore, and M.~Gross.
\newblock Iterative image warping.
\newblock {\em Comp. Graph. Forum (Proc. Eurographics)}, 31(2), 2012.

\bibitem{Buehler2001}
C.~Buehler, M.~Bosse, L.~McMillan, S.~Gortler, and M.~Cohen.
\newblock Unstructured lumigraph rendering.
\newblock In {\em Proc. SIGGRAPH}, 2001.

\bibitem{Chaurasia2013}
G.~Chaurasia, S.~Duchene, O.~Sorkine-Hornung, and G.~Drettakis.
\newblock Depth synthesis and local warps for plausible image-based navigation.
\newblock {\em ACM Trans. Graph.}, 32(3), 2013.

\bibitem{dabala2016lightfield}
{\L}.~Daba{\l}a, M.~Ziegler, P.~Didyk, F.~Zilly, J.~Keinert, K.~Myszkowski,
  H.-P. Seidel, P.~Rokita, and T.~Ritschel.
\newblock {Efficient Multi-image Correspondences for On-line Light Field Video
  Processing}.
\newblock {\em Comp. Graph. Forum (Proc. Pacific Graphics)}, 2016.

\bibitem{DidykERMS2010}
P.~Didyk, T.~Ritschel, E.~Eisemann, K.~Myszkowski, and H.-P. Seidel.
\newblock Adaptive image-space stereo view synthesis.
\newblock In {\em Proc. VMV}, 2010.

\bibitem{dosovitskiy2015learning}
A.~Dosovitskiy, J.~Tobias~Springenberg, and T.~Brox.
\newblock Learning to generate chairs with convolutional neural networks.
\newblock In {\em CVPR}, 2015.

\bibitem{DuDDHM2013}
S.-P. Du, P.~Didyk, F.~Durand, S.-M. Hu, and W.~Matusik.
\newblock Improving visual quality of view transitions in automultiscopic
  displays.
\newblock {\em ACM Trans. Graph. (Proc. SIGGRAPH)}, 33(6), 2014.

\bibitem{engel2017neural}
J.~Engel, C.~Resnick, A.~Roberts, S.~Dieleman, M.~Norouzi, D.~Eck, and
  K.~Simonyan.
\newblock Neural audio synthesis of musical notes with wavenet autoencoders.
\newblock In {\em JMLR}, 2017.

\bibitem{flynn2019deepview}
J.~Flynn, M.~Broxton, P.~Debevec, M.~DuVall, G.~Fyffe, R.~Overbeck, N.~Snavely,
  and R.~Tucker.
\newblock Deepview: View synthesis with learned gradient descent.
\newblock In {\em CVPR}, 2019.

\bibitem{flynn2016deepstereo}
J.~Flynn, I.~Neulander, J.~Philbin, and N.~Snavely.
\newblock Deepstereo: Learning to predict new views from the world's imagery.
\newblock In {\em CVPR}, 2016.

\bibitem{ohad2019dubbing}
O.~Fried and M.~Agrawala.
\newblock Puppet dubbing.
\newblock In {\em Proc. EGSR}, 2019.

\bibitem{Fuchs2007}
M.~Fuchs, V.~Blanz, H.~P. Lensch, and H.-P. Seidel.
\newblock Adaptive sampling of reflectance fields.
\newblock {\em ACM Trans. Graph.}, 26(2), 2007.

\bibitem{monodepth17}
C.~Godard, O.~{Mac Aodha}, and G.~J. Brostow.
\newblock Unsupervised monocular depth estimation with left-right consistency.
\newblock In {\em CVPR}, 2017.

\bibitem{goodfellow2014generative}
I.~Goodfellow, J.~Pouget-Abadie, M.~Mirza, B.~Xu, D.~Warde-Farley, S.~Ozair,
  A.~Courville, and Y.~Bengio.
\newblock Generative adversarial nets.
\newblock In {\em Proc. NIPS}, 2014.

\bibitem{gortler1996lumigraph}
S.~J. Gortler, R.~Grzeszczuk, R.~Szeliski, and M.~F. Cohen.
\newblock The lumigraph.
\newblock In {\em SIGGRAPH}, 1996.

\bibitem{hedman2018blending}
P.~Hedman, J.~Philip, T.~Price, J.-M. Frahm, G.~Drettakis, and G.~J. Brostow.
\newblock Deep blending for free-viewpoint image-based rendering.
\newblock {\em ACM Trans. Graph. (Proc. SIGGRAPH)}, 37(6), 2018.

\bibitem{hedman2016scalablerendering}
P.~Hedman, T.~Ritschel, G.~Drettakis, and G.~Brostow.
\newblock Scalable inside-out image-based rendering.
\newblock {\em ACM Trans. Graph. (Proc. SIGGRAPH Asia)}, 35(6), 2016.

\bibitem{hinton2006reducing}
G.~E. Hinton and R.~R. Salakhutdinov.
\newblock Reducing the dimensionality of data with neural networks.
\newblock {\em Science}, 313(5786), 2006.

\bibitem{jaderberg2015spatial}
M.~Jaderberg, K.~Simonyan, A.~Zisserman, et~al.
\newblock Spatial transformer networks.
\newblock In {\em Proc. NIPS}, 2015.

\bibitem{Kalantari2016}
N.~K. Kalantari, T.-C. Wang, and R.~Ramamoorthi.
\newblock Learning-based view synthesis for light field cameras.
\newblock {\em ACM Trans. Graph. (Proc. SIGGRAPH Asia)}, 35(6), 2016.

\bibitem{Kellnhofer2017}
P.~Kellnhofer, P.~Didyk, S.-P. Wang, P.~Sitthi-Amorn, W.~Freeman, F.~Durand,
  and W.~Matusik.
\newblock {3DTV} at home: Eulerian-lagrangian stereo-to-multiview conversion.
\newblock {\em ACM Trans. Graph. (Proc. SIGGRAPH)}, 36(4), 2017.

\bibitem{kingma2013auto}
D.~P. Kingma and M.~Welling.
\newblock Auto-encoding variational bayes.
\newblock In {\em Proc. ICLR}, 2013.

\bibitem{leimkuehler2017microwarping}
T.~Leimk\"uhler, H.-P. Seidel, and T.~Ritschel.
\newblock Minimal warping: Planning incremental novel-view synthesis.
\newblock {\em Comp. Graph. Form (Proc. EGSR)}, 36(4), 2017.

\bibitem{levoy1996light}
M.~Levoy and P.~Hanrahan.
\newblock Light field rendering.
\newblock In {\em SIGGRAPH}, 1996.

\bibitem{liu2018coordconv}
R.~Liu, J.~Lehman, P.~Molino, F.~P. Such, E.~Frank, A.~Sergeev, and
  J.~Yosinski.
\newblock An intriguing failing of convolutional neural networks and the
  coordconv solution.
\newblock In {\em Proc. NIPS}, 2018.

\bibitem{liu2019softras}
S.~Liu, T.~Li, W.~Chen, and H.~Li.
\newblock Soft rasterizer: A differentiable renderer for image-based {3D}
  reasoning.
\newblock {\em ICCV}, 2019.

\bibitem{liu2017video}
Z.~Liu, R.~A. Yeh, X.~Tang, Y.~Liu, and A.~Agarwala.
\newblock Video frame synthesis using deep voxel flow.
\newblock In {\em Proc. ICCV}, 2017.

\bibitem{lombardi2019volumes}
S.~Lombardi, T.~Simon, J.~Saragih, G.~Schwartz, A.~Lehrmann, and Y.~Sheikh.
\newblock Neural volumes: Learning dynamic renderable volumes from images.
\newblock {\em ACM Trans. Graph. (Proc. SIGGRAPH)}, 38(4), 2019.

\bibitem{mark1997post}
W.~R. Mark, L.~McMillan, and G.~Bishop.
\newblock Post-rendering {3D} warping.
\newblock In {\em Proc. i3D}, 1997.

\bibitem{maximov2019dam}
M.~Maximov, L.~Leal-Taixé, M.~Fritz, and T.~Ritschel.
\newblock Deep appearance maps.
\newblock In {\em Proc. ICCV}, 2019.

\bibitem{Mildenhall2019}
B.~Mildenhall, P.~P. Srinivasan, R.~Ortiz-Cayon, N.~K. Kalantari,
  R.~Ramamoorthi, R.~Ng, and A.~Kar.
\newblock Local light field fusion: Practical view synthesis with prescriptive
  sampling guidelines.
\newblock {\em ACM Trans. Graph. (Proc. SIGGRAPH)}, 38(4), 2019.

\bibitem{nguyen2018rendernet}
T.~{Nguyen Phuoc}, C.~Li, S.~Balaban, and Y.~Yang.
\newblock Rendernet: A deep convolutional network for differentiable rendering
  from 3d shapes.
\newblock 2018.

\bibitem{Soft3DReconstruction}
E.~Penner and L.~Zhang.
\newblock Soft {3D} reconstruction for view synthesis.
\newblock {\em ACM Trans. Graph. (Proc. SIGGRAPH Asia)}, 36(6), 2017.

\bibitem{petersen2019pix2vec}
F.~Petersen, A.~H. Bermano, O.~Deussen, and D.~Cohen{-}Or.
\newblock Pix2vex: Image-to-geometry reconstruction using a smooth
  differentiable renderer.
\newblock {\em Arxiv abs/1903.11149}, 2019.

\bibitem{radford2015unsupervised}
A.~Radford, L.~Metz, and S.~Chintala.
\newblock Unsupervised representation learning with deep convolutional
  generative adversarial networks.
\newblock {\em Arxiv abs/1511.06434}, 2015.

\bibitem{rainer2019btf}
G.~Rainer, W.~Jakob, A.~Ghosh, and T.~Weyrich.
\newblock Neural btf compression and interpolation.
\newblock {\em Comp. Graph. Forum (Proc. Eurographics)}, 38(2), 2019.

\bibitem{reed2015deep}
S.~E. Reed, Y.~Zhang, Y.~Zhang, and H.~Lee.
\newblock Deep visual analogy-making.
\newblock In {\em NIPS}, 2015.

\bibitem{ren2015image}
P.~Ren, Y.~Dong, S.~Lin, X.~Tong, and B.~Guo.
\newblock Image based relighting using neural networks.
\newblock {\em ACM Trans. Graph. (Proc. SIGGRAPH)}, 34(4), 2015.

\bibitem{Sabater2017}
N.~Sabater, G.~Boisson, B.~Vandame, P.~Kerbiriou, F.~Babon, M.~Hog,
  T.~Langlois, R.~Gendrot, O.~Bureller, A.~Schubert, and V.~Allie.
\newblock Dataset and pipeline for multi-view light-field video.
\newblock In {\em CVPR Workshops}, 2017.

\bibitem{sitzmann2019deepvoxels}
V.~Sitzmann, J.~Thies, F.~Heide, M.~Nie{\ss}ner, G.~Wetzstein, and
  M.~Zollh{\"o}fer.
\newblock Deepvoxels: Learning persistent 3d feature embeddings.
\newblock In {\em CVPR}, 2019.

\bibitem{Srinivasan2017}
P.~P. Srinivasan, T.~Wang, A.~Sreelal, R.~Ramamoorthi, and R.~Ng.
\newblock Learning to synthesize a {4D} rgbd light field from a single image.
\newblock {\em ICCV}, 2017.

\bibitem{Sun2018PWC-Net}
D.~Sun, X.~Yang, M.-Y. Liu, and J.~Kautz.
\newblock {PWC-Net}: {CNNs} for optical flow using pyramid, warping, and cost
  volume.
\newblock In {\em CVPR}, 2018.

\bibitem{vondrick2016generating}
C.~Vondrick, H.~Pirsiavash, and A.~Torralba.
\newblock Generating videos with scene dynamics.
\newblock In {\em Proc. NIPS}, 2016.

\bibitem{wang2018vid2vid}
T.-C. Wang, M.-Y. Liu, J.-Y. Zhu, G.~Liu, A.~Tao, J.~Kautz, and B.~Catanzaro.
\newblock Video-to-video synthesis.
\newblock In {\em NeurIPS}, 2018.

\bibitem{WG13_tpami}
S.~Wanner and B.~Goldluecke.
\newblock Variational light field analysis for disparity estimation and
  super-resolution.
\newblock {\em PAMI}, 36(3), 2014.

\bibitem{white2016sampling}
T.~White.
\newblock Sampling generative networks.
\newblock {\em Arxiv 1609.04468}, 2016.

\bibitem{Xu2019}
Z.~Xu, S.~Bi, K.~Sunkavalli, S.~Hadap, H.~Su, and R.~Ramamoorthi.
\newblock Deep view synthesis from sparse photometric images.
\newblock {\em ACM Trans. Graph. (Proc. SIGGRAPH)}, 38(4), 2019.

\bibitem{YangNehab2009}
L.~Yang, D.~Nehab, P.~V. Sander, P.~Sitthi-amorn, J.~Lawrence, and H.~Hoppe.
\newblock Amortized supersampling.
\newblock {\em ACM Trans. Graph.}, 28(5), 2009.

\bibitem{Zhou2018StereoMagnification}
T.~Zhou, R.~Tucker, J.~Flynn, G.~Fyffe, and N.~Snavely.
\newblock Stereo magnification: Learning view synthesis using multiplane
  images.
\newblock {\em ACM Trans. Graph. (Proc. SIGGRAPH)}, 37(4), 2018.

\bibitem{zhou2016appearance}
T.~Zhou, S.~Tulsiani, W.~Sun, J.~Malik, and A.~A. Efros.
\newblock View synthesis by appearance flow.
\newblock In {\em ECCV}, 2016.

\end{thebibliography}

\end{document}